# Theoretical Design of Solid Electrolytes with Superb Ionic Conductivity: Alloying Effect on $Li^+$ Transportation in Cubic $Li_6PA_5X$ Chalcogenides

Zhuo Wang, Min Jie Xuan, Hong Jie Xu, and Guosheng Shao*

State Center for Designer Low-carbon & Environmental Materials, Zhengzhou University, 100 Kexue Avenue, Zhengzhou 450001, China

Zhengzhou Materials Genome Institute, Zhongyuanzhigu, Xingyang 450100, China

*Corresponding author email: gsshao@zzu.edu.cn

**Abstract:** It is of great importance to develop inorganic solid electrolytes with high ionic conductivity, thus enabling solid state Li-ion batteries to address the notorious safety issue about the current technology due to use of highly flammable liquid organic electrolytes. On the basis of systematic first principles modelling, we have formulated new inorganic electrolytes with ultra-low activation energies for long-distance diffusion of $Li^+$ ions, through alloying in the cubic $Li_6PA_5X$ chalcogenides (chalcogen A; halogen X). We find that the long-distance transportation of $Li^+$ is dictated by inter-octahedral diffusion, as the activation energy for $Li^+$ to migrate over a $Li_6A$ octahedron is minimal. The inter-octahedral diffusion barrier for $Li^+$ is largely dependent on the interaction with chalcogen anions in the compound. Radical reduction of diffusion barrier for $Li^+$ ions can be realized through isovalent substitution of S using elements of lower electronegativity, together with smaller halogen ions on the X site.

**Keywords:** Materials formulation; First principle modelling; Superfast solid ionic conductor; Solid electrolytes; Li-ion battery

## Introduction

Li-ion batteries have been widely exploited as an advanced energy storage technology for various applications such as portable electronics and electric vehicles.[1, 2] Current commercial lithium-ion batteries, however, are in exposure to potential safety hazards due to the flammable organic liquid electrolytes being used. Indeed, most recent major incidents of Li-ion battery fires were caused by ignition of the liquid electrolyte from overcharging or abusing operations.[3,4] Great efforts have therefore been directed towards developing fundamentally safe technological solutions using inorganic solid electrolytes.

Research endeavors in developing solid electrolytes was initially focused on $Li^+$ conducting oxides. For example, $Li_{1.3}M_{0.3}Ti_{1.7}(PO_4)_3$ (M=Al and Sc) based on lithium titanium phosphates were prepared,[6] with maximum conductivity at 298K being only 0.7 $mScm^{-1}$. The $Li_7La_3Zr_2O_{12}$ had an upper limit of conductivity of 0.774 $mScm^{-1}$ at 25 ℃,[7] and the $Li_{2.88}PO_{3.73}N_{0.14}$ showed an average conductivity of $2.3(\pm0.7)\times10^{-3}$ $mScm^{-1}$ at 25 ℃ with an average activation energy $E_a$ about 0.55(±0.02) eV.[8] Overall, these oxide based Li ion conductors exhibited ionic conductivities ranging from $10^{-3}$ to 1 $mScm^{-1}$ at room temperature, with the activation energy being in the range from 0.3 to 0.6 eV, which are significantly inferior to the conductivity of current organic liquid electrolytes for Li-ion batteries (up to 10 $mScm^{-1}$).[9]

Recently, great advancements have been made in solid electrolytes with compositions around $Li_{4-x}Ge_{1-x}P_xS_4$ or the LGPS family,[10] which are based on sulfides.[11-12] The near equivalent amounts of S and Li are key for large capacity of $Li^+$ in the electrolytes, while introducing other elements such as Ge and P help improve the ionic conductivity and lower the activation energy for diffusion of $Li^+$ ions. For example, with x = 0.75, the ionic conductivity reached 2.2 $mScm^{-1}$ at room temperature.[10] A latest member of the LGPS family, $Li_{10}GeP_2S_{12}$, is constructed by $PS_4$ and $GeP_4$ tetrahedrons without the di-tetrahedral units. The tetragonal $Li_{10}GeP_2S_{12}$ phase is among the best solid $Li^+$ conducting electrolytes, exhibiting an extremely high bulk conductivity of over 10 $mScm^{-1}$ at room temperature.[13,14] The efficient conduction pathway for $Li^+$ diffusion in $Li_{10}GeP_2S_{12}$ is, however, only one-dimensional, with an activation energy barrier for diffusion being 0.22-0.25 eV.[13,14] Replacing the expensive Ge led to a new formulation from the LGPS family, with a $Li_{9.54}Si_{1.74}P_{1.44}S_{11.7}Cl_{0.3}$ being developed,[15,16] which had a very high ionic conductivity of about 25 mScm$^{-1}$ at room temperature, even exceeding that of organic liquid electrolytes. However, it was found that this Ge-free electrolyte was not electrochemically stable with the lithium anode. On the other hand, the other more stable Ge-free electrolyte in the same report, $Li_{9.6}P_3S_{12}$, had a much lower ionic conductivity of 1.2 mScm$^{-1}$.[16]

Theoretical simulation revealed that in the $Li_{10}GeP_2S_{12}$ system, a body-centered cubic (bcc) anion framework of S can offer channels of low activation barrier and high ionic conductivity.[17] Although the expensive Ge in the LGPS could be substituted by cheaper Sn or Si, it has limited impact on stability and conductivity.[18] Replacing S with larger diameter Se had rather limited effect on improving $Li^+$ conductivity, suggesting that there is a critical diffusion channel size in the LGPS system, beyond which $Li^+$ diffusion cannot be improved further.[18]

In spite of the recent landmark advancement in achieving high $Li^+$ conductivity rivalling that of liquid organic counterparts in sulfide based solid electrolytes; it is yet highly desirable to develop solid electrolytes with both superb $Li^+$ ion conductivity and high electrochemical stability to enable direct use of Li anode in full batteries. It is envisaged that competitive solid electrolytes should possess the following characteristics: (a) Having large $Li^+$ capacity to permit a big $Li^+$ concentration difference between the electrolyte and anode, which is beneficial for overcoming interfacial resistance in the $Li^+$ transport process;[5] (b) Providing efficient $Li^+$ transport pathways with low activation energy barrier; (c) Being chemically stable and electrochemically compatible with the Li anode; and (d) Of good mechanical stability over a wide temperature range for improved cycling performance.

A group of high $Li^+$ sulfides, $Li_6PS_5X$ with X being a halogen element, was reported to be electrochemically stable at least up to 5 V vs. $Li/Li^+$.[19,25] This group of materials are of the cubic argyrodite structure of the $Ag_8GeS_6$ phase (space group $F\bar{4}3m$), which is a moderate ionic conductor for $Ag^+$.[20] The experimentally reported $Li^+$ conductivity, however, was rather scattered even for the material with the same halogen element.[20,21] Such scatter in reported ionic conductivity could be attributed to difference in either processing conditions or resultant compositions.[20,22-25]

In this work, we attempt for a systematic theoretical approach in designing superionic $Li^+$

ion conductors, through isovalent alloying on the chalcogen (A) and halogen (X) sites in the cubic $Li_6PA_5X$ structure. We aim at offering fundamental guidance towards formulating a new family of excellent ionic conductors as promising alternatives to the tetragonal LGPS based materials, via addressing the issue of inadequate electrochemical compatibility with Li metal, while enabling superfast ionic transport with even lowered activation energy for $Li^+$ diffusion.

## Methods

Theoretical calculations are performed using the Vienna *ab initio* Simulation Package (VASP),[26,27] with the ionic potentials including the effect of core electrons being described by the projector augmented wave (PAW) method.[28,29] In this work, the Perdew−Burke−Ernzerhof (PBE) GGA exchange−correlation (XC) functionals[30,31] are used to relax the structural configurations in the $Li_6PA_5X$ family, allowing isovalent substitution of the A and X sites. Although standard DFT calculations (GGA) are sometimes inadequate to describe the localization of excited electrons, particularly in correlated metal oxides, they are largely suitable for the simulation of structural-energetic properties at the ground state, with the pronounced advantage of low computing cost. For the geometric relaxation of the structures, summation over the Brillouin Zone (BZ) is performed with 3×3×3 and 5×5×5 Monkhorst−Pack k-point mesh for the conventional and primitive cells respectively. A plane-wave energy cutoff of 500 eV is used in all calculations. All structures are geometrically relaxed until the total force on each ion was reduced below 0.01 eV/Å. For the calculations of electronic energy band structures, we use the HSE06 functional to predict more accurate values of band gaps.[32,33] We employ a convergence criterion of $10^{-6}$ eV for electronic self-consistent cycle.

The climb image nudged elastic band (CI-NEB) method with the Limited-memory Broyden-Fletcher-Goldfarb-Shanno (LBFGS) optimizer[34,35] has been used to search the $Li^+$ diffusion pathways in the electrolytes of interest. The initial and final configurations are obtained after full structural relaxation. The number of inserted images used in the CI-NEB calculations depend on the reaction coordinates between the initial and final configurations. Ab initio molecular dynamics (AIMD) is then carried out to study the ionic transport behaviour of Li ions at elevated temperatures, thus providing further insights through statistic processes. For the sake of tractable AIMD tasks, the simulation runs are performed on the conventional cells of $Li_6PA_5X$ containing 52 atoms, with 2 fs as time step in the NVT ensembles together with a Nosé–Hoover thermostat.[22] Each AIMD run lasts for 80 ps after a 10 ps pre-equilibrium run. In order to shorten the simulation time, elevated temperatures from 600 to 1200 K are applied to speed up the ion-hopping process.

The universal structure predictor (USPEX)[36,37] based on a materials genome algorithm has been employed to predict *globally stable and metastable structures with variable compositions*. For each composition, a population of 200 possible structures is created randomly with varied symmetries in the first generation. When the full structure relaxation is reached, the most stable and metastable structures, through the comparison of enthalpy of formation, will be inherited into the next generation. Afterwards, each subsequent generation will be created through heredity, with lattice mutation and permutation operators being applied and assessed energetically for the selection of a population of 60 for the subsequent run. USPEX will continue screening the structures until the most stable configuration stays unchanged for further 20 generations to safeguard global equilibrium.

The phonon frequency spectrum of a theoretically predicted structure is used for examining its dynamical stability. The supercell method in the PHONOPY package[38,39] is employed to perform the relevant frozen-phonon calculations with harmonic approximation. The supercells containing $2\times2\times2$ primitive cells (104 atoms) of the relaxed $Li_6PA_5X$ structures are used for the phonon calculations. The stability criterion is that the amplitude of imaginary frequency is less than 0.3 THz,[40,41] to accommodate numerical errors in phonon calculations.

## Results and Discussion

### *Global searching for low energy configurations:*

The symmetry of $Li_6PS_5X$ belongs to the face-centered cubic $F\bar{4}3m$ space group at room temperature,[19-21] which is constructed by regular octahedral $Li_6S$ and tetrahedral $PS_4$ units as shown in Fig. 1, for the $Li_6PS_5I$ as an example (conventional cell containing 52 atoms). The S atoms are located either in the center of the octahedral $Li_6S$ units or at the corners of the tetrahedral $PS_4$ units. Indeed, the chemical environments around these two types of S are quite different. In addition, the halogen species such as the iodine atoms (I) locate in the interstitial site between the tetrahedral $PS_4$ units.

We now investigate the effect of isovalent substitution of either the chalcogen or halogen sites, on the thermodynamic stability and associated physical properties of the $Li_6PS_5I$ phase. The stable crystal structure of each alloyed phase is identified from varied crystallographic configurations on the basis of global energetic minimization, using the USPEX search engine[36,37] for the specified chemical constitution, such as $Li_6PSO_4I$, $Li_6PS_5I$, $Li_6PSeS_4I$, $Li_6PTeS_4I$, $Li_6PSe_5I$, $Li_6PTe_5I$, $Li_6PS_5Br$, $Li_6PS_5Cl$, $Li_6PTeS_4Cl$, and $Li_6PTe_5Cl$. Such a theoretical approach is particularly useful when little is known about phase structures in a new materials system to be formulated, so that potential phase structures can be predicted with associated properties evaluated.

On the basis of the global structure searching using the USPEX approach, one can generalize some fundamental principles for the formation of structural configurations in each alloy. Firstly, $P^{+5}$ cations prefer to bond with four bivalent chalcogen anions that have larger electronegativity (O>S>Se>Te>P) to build a regular tetrahedral P-centered unit, while leaving the other chalcogen anion of weaker electronegativity to be encaged within six Li atoms to form the Li-chalcogen octahedrons. For example, in $Li_6PTeS_4I$, S stays on the corners of $PS_4$ while Te is located in the center of $Li_6Te$. In the same way, there are $PO_4$ and $Li_6S$ in the $Li_6PSO_4I$ phase, and $PS_4$ and $Li_6Se$ in $Li_6PSeS_4I$ and so on. Secondly, halogen atoms tend to locate between polyhedrons without significant contribution to chemical bonding, such that the enthalpy of formation for $Li_6PS_5X$ (X=I, Br, Cl) are rather similar, indicating their weaker electrostatic interaction with the cations. Moreover, their effect on the lattice parameter is rather trivial. For example, the lattice parameter of 10.31 Å for $Li_6PS_5I$ is only marginally bigger than those for $Li_6PS_5Br$ (10.29Å) and $Li_6PS_5Cl$ (10.27Å). This is much less effective than isovalent substitution of the S sites. For example, in the $Li_6PA_5I$ group, the size of cell changes from 8.84 to 11.78 Å, in the sequence of $Li_6PTe_5I > Li_6PSe_5I > Li_6PTeS_4I > Li_6PSeS_4I > Li_6PS_5I > Li_6PSO_4I$. In the $Li_6PA_5Cl$ group, the same trend for lattice parameter also holds: $Li_6PTe_5Cl > Li_6PTeS_4Cl > Li_6PS_5Cl$. The significant change in lattice parameter due to isovalent substitution of S is in accord with the difference in the ionic diameters of the elements in the S-group. Such significant effect in changing the lattice parameter is, in turn, closely correlated to their effect in changing the chemical bonding strength in the material.

***Structure stability and electrochemical compatibility with Li anode：***

The high symmetry structure becomes most stable when 80% of S is replaced by O. Actually, isovalent substitution of S, except for O, in the $Li_6PA_5X$ phase induces little change in the thermodynamic stability with respect to that of the fully relaxed P1 phase for the $Li_6PA_5X$ family. Tables 1 and 2 summarize the change in lattice parameter and symmetry due to structural relaxation at 0 K, and the corresponding change in total energy is presented in Fig. 2. The difference in the enthalpy of formation due to the loss of symmetry in the relaxed structures are insignificant, being 0.02, 0.026, 0.038, 0.038 and 0.07eV for $Li_6PS_5I$, $Li_6PSeS_4I$, $Li_6PTeS_4I$, $Li_6PSe_5I$, and $Li_6PTe_5I$, correspondingly. For phases containing a halogen element other than I, the differences are 0.04, 0.05, 0.08 and 0.1 eV per atom for $Li_6PS_5Br$, $Li_6PS_5Cl$, $Li_6PTeS_4Cl$, and $Li_6PTe_5Cl$ correspondingly. The change in the thermodynamic stability due to such isovalent replacement is in line with the electronegativity and associated ionic diameter of the substituting element, so that O replacement of S lowers the energy of formation value due to its greater electronegativity than that for S, thus leading to stronger binding between the anions and $Li^+$ ions. Replacement of S by Te, on the other hand, leads to weakened binding, as the electronegativity of Te is lower than that of S.

The crystal structures of the $Li_6PA_5I$ alloys (A=S, O, Se, Te) are presented in Fig. 3, where both conventional and primitive cells are shown for the face-centered cubic phase, with the primitive cell of the fully relaxed P1 structure shown at the lower right panel of each alloy. Taking from Fig. 3(a) for example, the basic octahedral and tetrahedral units are fundamental in the primitive cells for both the face centered phase and corresponding relaxed stable structure. Actually, the $PS_4$ unit still maintains the regular tetrahedral configuration in the P1 structure, though the $Li_6S$ unit is slightly distorted from the regular octahedral geometry, which is largely the origin for the loss of crystal symmetry. The maintenance of the regular tetrahedral configuration of the $PS_4$ unit is consistent with their bigger contribution to bonding. It is worth emphasizing that *the fully relaxed structure is only slightly off the high symmetry* space group of the cubic phase at 0 K, with deviation of atomic coordinates being well within 10% from their high-symmetry positions. Such full structural relaxation enables minimization of elastic energy, which is essential for dynamical stability of solid phases.

Now, let us check the dynamical stability of the fully relaxed structures through the characteristics of their phonon band structures (Dynamically stable phases are more likely to exist in nature owing to safeguarded mechanical stability). We chose each supercell made of $2\times2\times2$ primitive cells with 104 atoms for the phonon calculation. Phonon band structures of $Li_6PS_5I$, $Li_6PSe_5I$, $Li_6PTe_5I$, $Li_6PSO_4I$, $Li_6PSeS_4I$, $Li_6PTeS_4I$, $Li_6PS_5Cl$, $Li_6PTeS_4Cl$, and $Li_6PTe_5Cl$ in the low-frequency regions are displayed in Fig. 4, where the blue line marks zero frequency. The gray line stands for stability criterion of -0.3 THz in phonon calculation, which is caused by acceptable numerical errors.[40,41] If the amplitude of imaginary frequency is less than 0.3 THz, the system can be considered as dynamically stable. Existence of imaginary frequency below the -0.3 THz threshold is indicative of dynamical instability. It is seen from Fig. 4 that the -0.3 THz criterion is satisfied for all the fully relaxed structures, indicating that all of these structural configurations are mechanically stable.

Although at 0 K the face-centered cubic $Li_6PS_5Cl$ phase is metastable with respect to a linear combination of $Li_2S$, $Li_3PS_4$, and LiCl,[22, 42] it has been observed in experiments, e.g. $Li_6PS_5X$.[19-25] This indicates that the face-centered cubic $Li_6PS_5X$ phase could become entropically stabilized at elevated temperatures as a result of phonon excitations. We now examine the temperature effect on free energies for the alloys of interest, within the quasiharmonic approximation.[43,44] The Gibbs free energies for each compound can be calculated against other possible constituent candidates in the vicinity of each alloy,

$$\Delta G_{\mathrm{Li6PS5Cl}} = G_{\mathrm{Li6PS5Cl}} - G_{\mathrm{Li2S}} - G_{\mathrm{Li3PS4}} - G_{\mathrm{LiCl}},$$

$$\Delta G_{\mathrm{Li6PTeS4Cl}} = G_{\mathrm{Li6PTeS4Cl}} - G_{\mathrm{Li2Te}} - G_{\mathrm{Li3PS4}} - G_{\mathrm{LiCl}}$$

where all ground state stable constituent candidates, such as $Li_2S$, $Li_3PS_4$, LiCl, and $Li_2Te$, are also found to be dynamically/mechanically stable (exhibiting only real phonon frequencies in the phonon band structures).

The critical temperatures to enable negative Gibbs free energies of the cubic phases of $Li_6PS_5Cl$ and $Li_6PTeS_4Cl$ are indicated in Fig. 5. It suggests that the cubic $Li_6PS_5Cl$ will be stable above 613.9K, being close to its experimentally measured crystallization temperature of 330 $^0$C (603 K).[24] Similarly, the critical temperature for $Li_6PTeS_4Cl$ is at 614.2 K. The transition temperature of $Li_6PTeS_4Cl$ is almost the same as $Li_6PS_5Cl$, which suggests that the condition for synthetizing $Li_6PTeS_4Cl$ would be close to that of $Li_6PS_5Cl$.

The electrochemically compatible window with the anode Li metal can be examined according to interfacial reactions against Li uptake per formula unit (f.u.) of solid electrolyte.[22,45-46] Figure 6 (a) and (b) plots voltage above Li/$Li^+$ against the Li uptake per formula unit of $Li_6PS_5Cl$ and $Li_6PTeS_4Cl$ respectively. The solid electrolyte is to undergo reduction and uptake Li close to the alkali metal anode (Li rich side) at low voltage, while at high voltage (Li poor side), electrolyte will be oxidized to deplete Li. The intrinsic electrochemical window for the $Li_6PS_5Cl$ composition corresponds to a range of voltage (2.0−2.44 V), while the width of electrochemical window is between 1.97 and 2.29 V for $Li_6PTeS_4Cl$. Also, electronically insulating and ionically conducting phases of abundant $Li_2S$ are to be formed at higher voltages, which may potentially serve as good passivating interfacial phases to act as a barrier against further decomposition of electrolyte.

***Alloying effects on $Li^+$ diffusion via the vacancy mechanism:***

Diffusion of metallic species in condensed matter is usually dictated by the vacancy mechanism. Starting from the $Li_6PS_5I$ structure which has the largest lattice parameter for the $Li_6PS_5X$ phases with reported poor $Li^+$ transporting ability, let us try to find out the fundamental factors on restricting $Li^+$ diffusion. We use the CI-NEB method to quantify the activation energy barrier for migrating a $Li^+$ into a nearest vacancy site, using a structural model shown in Fig. 1, where arrows indicating most likely pathways. The red arrows in Fig. 1 shows equivalent diffusional paths within a single octahedral $Li_6S$ unit, or the *intra-octahedral* pathways, and the black arrow, on the other hand, corresponds to $Li^+$ migration from one octahedron to another, i.e. the *inter-octahedral* pathway.

*Intra-octahedral transport of $Li^+$*: We examine firstly the intra-octahedral diffusion over a single $Li_6S$ unit. The corresponding activation energy barrier for such intra-octahedral transport of $Li^+$ over a $Li_6S$ unit is as low as only 0.034 eV, Fig. 7(b). Since each S is surrounded by six $Li^+$ ions, the average attractive interaction from S to each Li is rather diluted, so that $Li^+$ experiences little electrostatic restriction from $S^{-2}$ and can then migrate easily. This is less than a sixth of that for the best LGPS electrolytes, and they impose little resistance to local $Li^+$ migration. As to the alloying effect due to isovalent substitution of the S site, we find that the energetic barrier for the migration of $Li^+$ ions over a $Li_6S$ octahedron is reversely correlated to the electronegativity of the chalcogen element, A, at the center of the $Li_6A$ octahedron, or at the corners of the $PA_4$ tetrahedron (A=O, S, Se), so that replacing four-fifth of the S atoms by O leads to significant increase in the $Li^+$ diffusion barrier over a $Li_6S$ unit (more than ten times from 0.034 to 0.38 eV), as shown in Fig. 7(b), albeit O only occupying the corner positions in the $PO_4$ tetrahedrons around each octahedron, as shown in Fig. 3(d). On the other hand, substituting one-fifth of S by Se can reduce the diffusion barrier instead (0.0288 eV), due to direct reduction of the electrostatic attraction between the central $Se^{-2}$ and the surrounding $Li^+$ ions in a $Li_6Se$ octahedron. The intra-octahedral diffusion barrier nearly disappears completely when the S sites are completely replaced by Se. In summary, the intra-octahedral diffusion barrier is ignorable as long as the A sites are not occupied by O.

However, long-distance diffusion of $Li^+$ ions cannot be realized without their jumping between neighboring $Li_6S$ octahedrons along $Li^+$ defect pathway that is indicated by a black arrow in Fig. 1. Fig. 7(a) shows the half-way transitional state for the inter-octahedral migration of a $Li^+$ ion, sitting within the triangular bottleneck made of one $I^-$ and two $S^{-2}$ anions in the $Li_6PS_5I$ structure. The inter-octahedral diffusion barrier in $Li_6PS_5I$ is 0.9 eV, Fig. 7(c), which suggests $Li_6PS_5I$ is a very poor $Li^+$ conductor.

*Alloying effect on inter-octahedral transport of $Li^+$*: We now move on to analyze the effect of S substitution on the inter-octahedral diffusion of $Li^+$ ions, which is the bottleneck for long-range transport of the $Li^+$ ions. Complete replacement of S with Se in $Li_6PS_5I$ reduces the barrier down to 0.6 eV. The inter-octahedral diffusion barrier can be decreased further by replacing 20% S with Te. This results in further reduction in diffusion barrier, being 0.52 eV in $Li_6PTeS_4I$.

The effect of isovalent substitution of S by elements of lower electronegativity and associated larger ionic diameter upon lowered diffusional barrier can be attributed to two factors: the increased lattice parameter offers larger channels for the transport of $Li^+$ ions, and the lowered electrostatic attraction between the polyhedral anions and the $Li^+$ ions.

The effects of the weaker halogen anions can be investigated through isovalent replacement of I by Br or Cl. We focus on their effect on the dominant diffusion barrier associated with the inter-octahedral diffusion of $Li^+$ ions. Replacing I with Cl or Br is found to reduce the inter-octahedral diffusion barrier significantly, from 0.9 eV for $Li_6PS_5I$ to 0.62 eV for $Li_6PS_5Br$, and 0.45eV for $Li_6PS_5Cl$, Fig.7 (d). The latter offers a promising basis for significant reduction of diffusion barriers through isovalent substitution of S with elements of lesser electronegativity such as Se and Te. Indeed, replacing one-fifth of the S sites with Te leads to over 42% reduction in the controlling inter-octahedral barriers for lithium diffusion, from 0.45 eV down to 0.26 eV. The activation barrier of 0.26 eV in $Li_6PTeS_4Cl$ is nearly equivalent to that for the fast $Li^+$ conductor $Li_{10}GeP_2S_{12}$ (0.22 – 0.25 eV),[13,14] with the latter being most effective for $Li^+$ diffusion only along the c-axis of its tetragonal structure.[5] The currently identified cubic electrolytes with three-dimensionally equivalent low activation barriers are highly promising in delivering superb $Li^+$ conductivity, as the activation energy barrier dictates the diffusion coefficient particularly at low temperatures.

The radically lowered inter-octahedral diffusion barrier for $Li^+$ ion is resultant from the combined effects of isovalent replacement of S with Te and the substitution of I with Cl. The effect of a smaller halogen anions than $I^-$ between the strongly bonded polyhedrons remains to be elaborated.

The halogen elements located in gaps between polyhedrons plays a rather trivial role in bonding, so that replacement of I with a smaller Br or Cl had rather insignificant effect on either bonding or lattice parameter. The major diffusional channel for $Li^+$ migration from one octahedron to another are shown in the charge density maps in Fig. 8(b-f), corresponding to a lithium ion being transported half way from one octahedron to the next (corresponding to the red lithium ion in Fig. 7a). The charge map cutting through a $Li_6S$ octahedron is displayed in Fig. 8(a), showing the equilibrium spacing between the S and Li (2.33 Å).

The charge maps for $Li_6PA_5X$ (A=S, O, Se, Te; X=Cl, Br, I) are compared in Fig. 8(b-d), where a marker for the gap between the Cl and S is shown as a reference scale. There is strong charge overlap between S and P, indicating strong covalent bonding. In contrast, the charge density between Li and S is rather low, being typical for ionic bonding. The ionic spacing between S and Li are seen to be correlated to the X-S spacing as well. It is noted that the I-S spacing is the largest due to the biggest ionic radius of I than those for Br and Cl, but the gap between I and S is the smallest instead. Likewise, the spacing between the $S^{-2}$ anion and $Li^+$ cation for $Li_6PS_5I$ is the smallest, indicating the strongest electrostatic interaction to hinder the migration of $Li^+$. Such outcome is consistent with the trivial effect of halogen elements on the lattice parameters and associated bonding strength in the $Li_6PS_5X$ phase, so that a halogen element of a bigger ionic diameter occupies more space in the diffusional channel, thus making it more difficult for a $Li^+$ to squeeze through. For example, while the atomic spacing between I and S is larger, the gap between them is smaller than that between S and Cl. The same effect holds true when S is replaced partly with Te. The significant reduction in the inter-octahedral diffusion barrier in $Li_6PTeS_4Cl$ is also consistent with the biggest S-Cl spacing of 4.45 Å, against 4.31 Å in $Li_6PS_5I$. Overall, $Li_6PTeS_4Cl$ has the largest effective size of $Li^+$ diffusion channel, largely owing to Cl occupying the smallest space, thus allowing the lowest activation energy of 0.26 eV for $Li^+$ transport.

***AIMD simulation*：**

While the CI-NEB method can be used to assess the activation barriers for ionic diffusion through path searching over a static energy landscape at 0 K, it is desirable to assess the temperature dependence of the ionic diffusion coefficient *D*. Recent work has demonstrated successful assessment of *D* by AIMD simulation at elevated temperatures.[22,47,48]

The diffusion coefficient is related to the average mean square displacement (MSD) from molecular dynamics (MD) runs at each temperature *T* over a period of time (t), $\langle[\Delta r(t)]^2\rangle$, as

$$D = \frac{1}{2dt}\langle[\Delta r(t)]^2\rangle,$$

where $d$ is the dimensionality factor. On the basis of the Arrhenius relation, $D = D_0 \exp(-\frac{E_a}{k_B T})$, one can derive $D_0$ and $E_a$ via plotting the logarithmic values of D against 1/T. Correction of the MSD from artefactual errors due to the periodic boundary condition is necessary for dependable AIMD evaluation of D, using the “unwrapped” trajectories to help achieve significant improvement in sampling statistics from limited simulation data typical for tractable AIMD runs.[49]

The ionic conductivity σ can then be derived using the Nernst−Einstein equation,[48]

$$\sigma = \frac{\rho z^2 F^2}{RT} D = \frac{\rho z^2 F^2}{RT} D_0 \exp(-\frac{E_a}{k_B T})$$

where $\rho$ is the molar density of diffusing alkali ions in the unit cell, z is the charge of alkali ions (+1for $Li^+$), F and R are the Faraday's constant and the gas constant, respectively, $D_0$ is a constant, $E_a$ is activation energy for diffusion, and $k_B$ is the Boltzmann constant. The ionic conductivity is thus dictated by the diffusion process.

The transport data for $Li^+$ are summarized in Table 3 to compare with CI-NEB data and data from literature. Initially AIMD is also carried out here to verify the current method using the well tested LPGS alloy as a benchmark (structural model containing 50 atoms), with data derived from the current work being in excellent agreement with experimental [13,14] and theoretical[47] data from the literature. On the other hand, the AIMD data from this work or literature[22] are of considerable discrepancy from data derived using the CI-NEB method. For example, in the case of the $Li_6PS_5Cl$ alloy, the AIMD simulation generated an activation energy for $Li^+$ somewhat bigger than that from that of the CI-NEB result (0.52 *vs.* 0.45 eV), which can be largely attributed to difference in the structural models as CI-NEB simulation, by convention, involves a vacancy in the supercell for path searching. It is highly encouraging that our AIMD result also show remarkably reduced activation energy for Li+ transport through substitution of 20% of S by Te, such that the activation energy for $Li_6PTeS_4Cl$ is down to only 0.175 eV, being about one third of that for the $Li_6PS_5Cl$.

It is worth noting that the reported experimental data[23,24] in the literature showed activation energies smaller than theoretical values. It was proposed in recent AIMD work on the $Li_6PS_5Cl$ phase that presence of some excess of $Li^+$ ions helps to promote their long-distance diffusion, thus leading to enhanced ionic conductivity.[22] Such off-stoichiometric effect is also investigated, while charge neutrality is maintained by replacing the halogen species Cl with chalcogen S, i.e. $Li_{6.25}PTeS_{4.25}Cl_{0.75}$ compound. It is found that such moderate off-stoichiometry drift helps lower the activation energy down to 0.168 eV. Overall, the considerably lowered activation energy owing to S replacement by Te enables highly enhanced ionic conductivity, such that the room temperature $Li^+$ conductivities in $Li_6PTeS_4Cl$ and $Li_{6.25}PTeS_{4.25}Cl_{0.75}$ are 91 and 126 mS $cm^{-1}$ respectively, about an order higher than that in the well-known LPGS alloys (10 – 12 mS $cm^{-1}$ at room temperature[13,14]) and alloys when the chalcogen site is fully occupied by S (14 mS $cm^{-1}$ in $Li_{6.25}PS_{5.25}Cl_{0.75}$)[22]. This indicates enormous potential for their usage as superfast ionic conductors for solid-state Li-ion batteries. The fact that partial substitution of S by Te alone leads to significant reduction of diffusion barriers is particularly encouraging for feasible application of the $Li_6PTeS_4Cl$ alloy, owing to ease in its practical synthesis owing to advantageous thermodynamic stability to hinder decomposition. Indeed, the current work has offered insightful guidance in successful synthesis of the $Li_6PTeS_4Cl$ in the same team.

Fig. 9 compares the temperature dependence of diffusion coefficients in alloys of interest to the current work, with comparison to other candidates. In addition to overall better performance, the lower activation energies owing to Te substitution of S is key for highly enhanced $Li^+$ transport at lower temperature. At room temperature, the diffusion coefficients for $Li_6PTeS_4Cl$ and $Li_{6.25}PTeS_{4.25}Cl_{0.75}$ are $0.7\times10^{-6}$ and $1\times10^{-6}$ $cm^2$ $s^{-1}$ respectively, almost one or two orders higher than that for LGPS or $Li_{6.25}PS_{5.25}Cl_{0.75}$.

***Bandgap:***

While a solid electrolyte needs to be highly conductive to $Li^+$ ions, it should be insulating to electrons to avoid internal discharging. Fig. 10 summarizes the density of states for the ionic conductors of interest to this work, using the HSE06 screened hybrid functional for improved accuracy in band structure calculation, by accounting for some non-local effect in the exchange-correlation functional.[16] It can be seen that all of the materials are of significant forbidden energy gaps between the valence band maximum (VBM) and conduction band minimum (CBM). Both the VBM and CBM are dominated by p electron states of anions in all compounds. Moreover, the forbidden gap is shown to be mainly determined by the type of bivalent nonmetal in the chalcogen site, so that the gap value is correlated to the overall bonding strength. The most strongly bonded O containing material has the largest forbidden gap than the S based materials, while S containing material has a larger band gap than the Se or Te-substituted materials. This can in turn be attributed to the higher partial density of states (DOS) next to the VBM from the halogen species than from the bivalent chalcogen species, such that a halogen species has only minor contribution to the overall bonding of the electrolytes. The DOS results predict that the band gap of $Li_6PTeS_4Cl$ is 2.5 eV, which would be adequate for preventing electrical conductivity so long as the alloy is not heavily doped either by impurity or native defects.

## Conclusions

New solid electrolytes with ultra-low diffusional barrier for $Li^+$ ions are designed through systematic first-principles modelling of alloying effect on the structures and properties of the $Li_6PA_5X$ (A=S, O, Se, Te; X=Cl, Br, I) based compounds with the cubic argyrodite structure.

Bonding in the $Li_6PS_5X$ is largely from the electrostatic interaction between the bivalent chalcogen anions and the $Li^+$ ions, so that replacing S by another chalcogen anion with lower electronegativity leads to weaker bonding and larger lattice.

The transport of $Li^+$ is dictated by inter-octahedral diffusion, which is largely dependent upon the interaction with anions in the compound, so that substitution of S by other chalcogen elements of lower electronegativity helps to significantly reduce the diffusional barrier well below that of the current-state-of-the-arts. Besides, the currently identified cubic solid electrolytes offer three-dimensionally equivalent low diffusion barriers for $Li^+$ ions, and are thus expected to have great potential as superionic conductors.

The $Li_6PTeS_4Cl$ identified in this work is of great potential as superb solid electrolyte for Li-ion batteries. Slight off-stoichiometry, with moderate excess of $Li^+$ and chalcogenide substitution of halide to maintain charge neutrality, is useful in further enhancement of the inter-octahedral diffusion of $Li^+$ ions.

## Acknowledgements

This work is supported in part by the 1000 Talents Program of China, the Zhengzhou Materials Genome Institute, the National Natural Science Foundation of China (Nos.51001091, 111174256, 91233101, 51602094, 11274100), and the Fundamental Research Program from the Ministry of Science and Technology of China (no. 2014CB931704). The authors also wish to thank Prof. Shyue Ping Ong and Dr. Iek-Heng Chu for useful discussions.

**TOC**

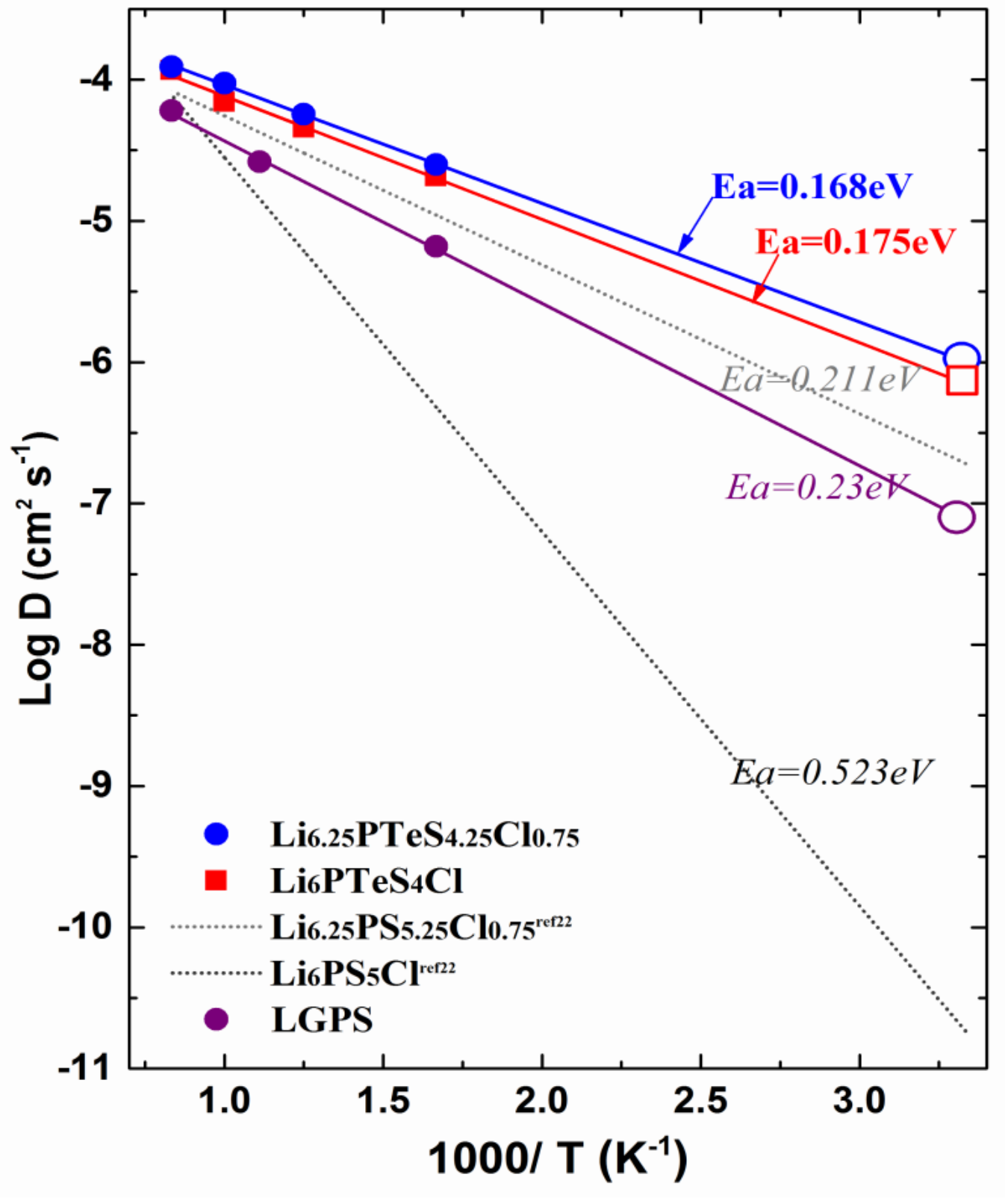

**Table 1** The lattice parameters and symmetry groups for $Li_6PS_5I$, $Li_6PSe_5I$, $Li_6PTe_5I$, $Li_6PSO_4I$, $Li_6PSeS_4I$, and $Li_6PTeS_4I$.

| | a | b | c (Å) | alpha | beta | gamma(°) | Symmetry |
|---|---|---|---|---|---|---|---|
| **$Li_6PS_5I$** | 10.31 | 10.31 | 10.31 | 90 | 90 | 90 | $F\bar{4}3M(216)$ |
| **--(primitive-cell)** | 7.29 | 7.29 | 7.29 | 60 | 60 | 60 | |
| **--(stable)** | 7.21 | 7.27 | 7.13 | 60.07 | 60.27 | 60.60 | *P*1(1) |
| **$Li_6PSe_5I$** | 10.88 | 10.88 | 10.88 | 90 | 90 | 90 | $F\bar{4}3M(216)$ |
| **--(primitive-cell)** | 7.69 | 7.69 | 7.69 | 60 | 60 | 60 | |
| **--( stable)** | 7.53 | 7.62 | 7.42 | 60.16 | 60.22 | 60.67 | *P*1(1) |
| **$Li_6PTe_5I$** | 11.78 | 11.78 | 11.78 | 90 | 90 | 90 | $F\bar{4}3M(216)$ |
| **--(primitive-cell)** | 8.33 | 8.33 | 8.33 | 60 | 60 | 60 | |
| **--( stable)** | 8.16 | 7.74 | 7.85 | 61.85 | 61.90 | 61.58 | *P1(1)* |
| **$Li_6PSO_4I$** | 8.84 | 8.84 | 8.84 | 90 | 90 | 90 | $F\bar{4}3M(216)$ |
| **--(primitive-cell)** | 6.25 | 6.25 | 6.25 | 60 | 60 | 60 | |
| **--( stable)** | -- | -- | -- | -- | -- | -- | -- |
| **$Li_6PSeS_4I$** | 10.46 | 10.46 | 10.46 | 90 | 90 | 90 | $F\bar{4}3M(216)$ |
| **--(primitive-cell)** | 7.40 | 7.40 | 7.40 | 60 | 60 | 60 | |
| **--( stable)** | 7.28 | 7.35 | 7.22 | 59.93 | 60.19 | 60.41 | *P1(1)* |
| **$Li_6PTeS_4I$** | 10.68 | 10.68 | 10.68 | 90 | 90 | 90 | $F\bar{4}3M(216)$ |
| **--(primitive-cell)** | 7.55 | 7.55 | 7.55 | 60 | 60 | 60 | |
| **--( stable)** | 7.44 | 7.39 | 7.39 | 60.02 | 60.4 | 60.88 | *P1(1)* |

**Table 2** The lattice parameters and symmetry groups for $Li_6PS_5Br$, $Li_6PS_5Cl$, $Li_6PTeS_4Cl$, and $Li_6PTe_5Cl$.

| | **a** | **b** | **c (Å)** | **alpha** | **beta** | **gamma(°)** | **Symmetry** |
|---|---|---|---|---|---|---|---|
| **$Li_6PS_5Br$** | 10.29 | 10.29 | 10.29 | 90 | 90 | 90 | $F\bar{4}3M$(216) |
| **--(primitive-cell)** | 7.27 | 7.27 | 7.27 | 60 | 60 | 60 | |
| **--(stable)** | 7.07 | 6.98 | 7.13 | 61.43 | 61.35 | 60.76 | *P*1(1) |
| | | | | | | | |
| **$Li_6PS_5Cl$** | 10.27 | 10.27 | 10.27 | 90 | 90 | 90 | $F\bar{4}3M$(216) |
| **--(primitive-cell)** | 7.26 | 7.26 | 7.26 | 60 | 60 | 60 | |
| **--( stable)** | 7.40 | 7.03 | 7.03 | 62.42 | 60.74 | 60.74 | *P1(1)* |
| | | | | | | | |
| **$Li_6PTeS_4Cl$** | 10.63 | 10.63 | 10.63 | 90 | 90 | 90 | $F\bar{4}3M$(216) |
| **--(primitive-cell)** | 7.51 | 7.51 | 7.51 | 60 | 60 | 60 | |
| **--( stable)** | 7.43 | 7.19 | 7.19 | 60.00 | 61.07 | 61.07 | *P1(1)* |
| | | | | | | | |
| **$Li_6PTe_5Cl$** | 11.76 | 11.76 | 11.76 | 90 | 90 | 90 | $F\bar{4}3M$(216) |
| **--(primitive-cell)** | 8.32 | 8.32 | 8.32 | 60 | 60 | 60 | |
| **--( stable)** | 8.17 | 7.72 | 7.53 | 62.72 | 61.92 | 63.35 | *P1(1)* |

**Table 3** Summary of data for ionic diffusion and conductivity, with reference to data from literature. Data from this work is emboldened

| System | Ea (eV) | $\sigma_{300}$ (mS cm$^{-1}$) |
|---|---|---|
| *$Li_6PS_5Cl$* | *0.38*[exp23] | *1.9*[exp23] |
| | *0.29*[exp24] | *1.1*[exp24] |
| | *0.524*[AIMD22] | *2.0*[AIMD22] |
| | **0.45**[CINEB] | |
| *$Li_{6.25}PS_{5.25}Cl_{0.75}$* | *0.211*[AIMD22] | *14*[AIMD22] |
| **$Li_6PTeS_4Cl$** | **0.26**[CINEB] | |
| | **0.175**[AIMD] | **91**[AIMD] |
| **$Li_{6.25}PTeS_{4.25}Cl_{0.75}$** | **0.168**[AIMD] | **126**[AIMD] |
| **LGPS** | *0.22*[exp13,14] | *10~12*[exp13,14] |
| | *0.21*[AIMD47] | *9*[AIMD47] |
| | **0.23**[AIMD] | **9.7**[AIMD] |

**Fig. 1** Structure of $Li_6PS_4I$, with an inset on the right showing the regular octahedron $Li_6S$ and tetrahedron $PS_4$ units. The arrows indicate key diffusion paths for $Li^+$: the intra-octahedral (red) and inter-octahedral (dark) paths for a $Li^+$ migrating to a vacancy.

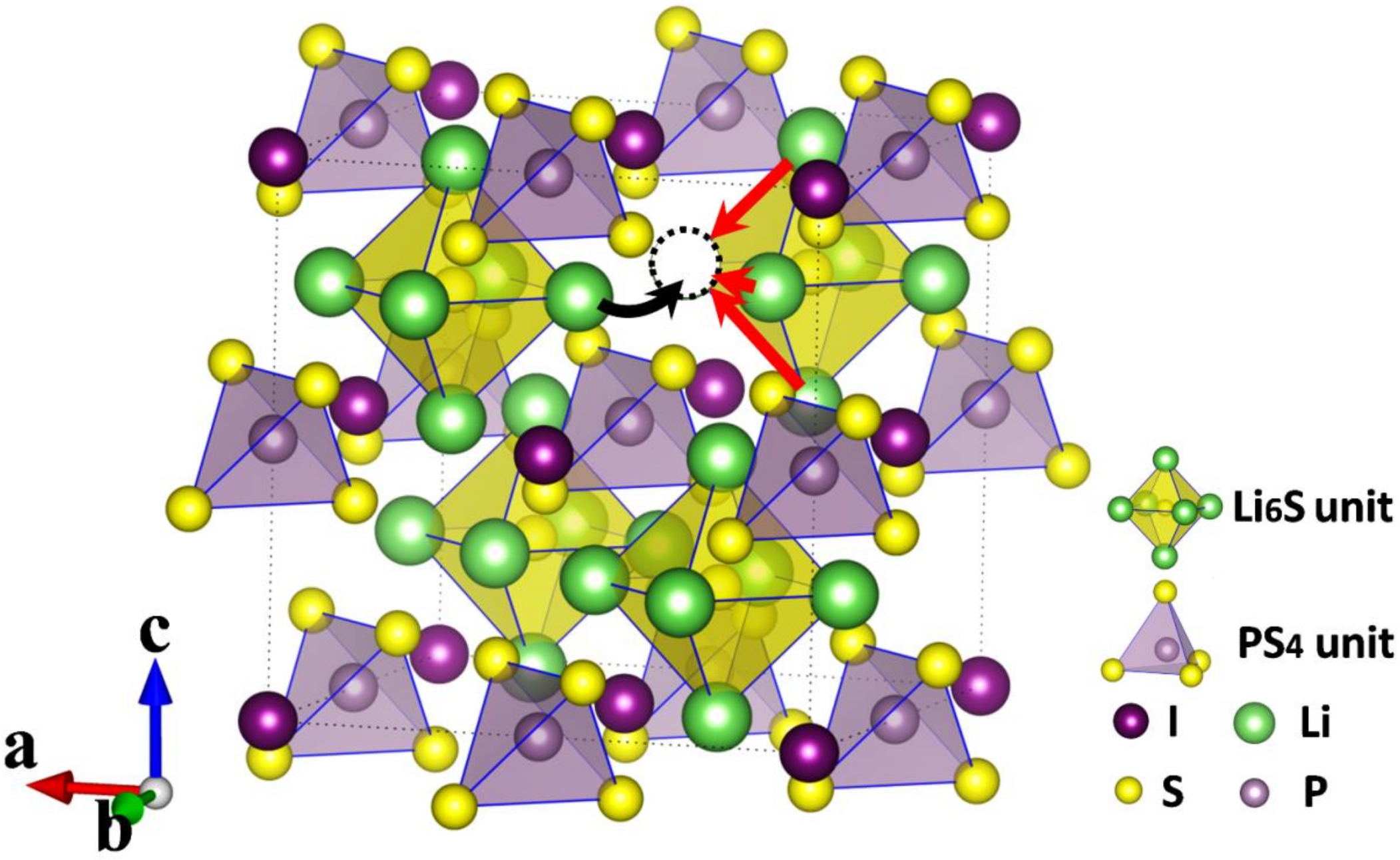

**Fig. 2** The global USPEX searched energies for the fully relaxed P1 (gray) and metastable cubic (yellow) structures in various alloys.

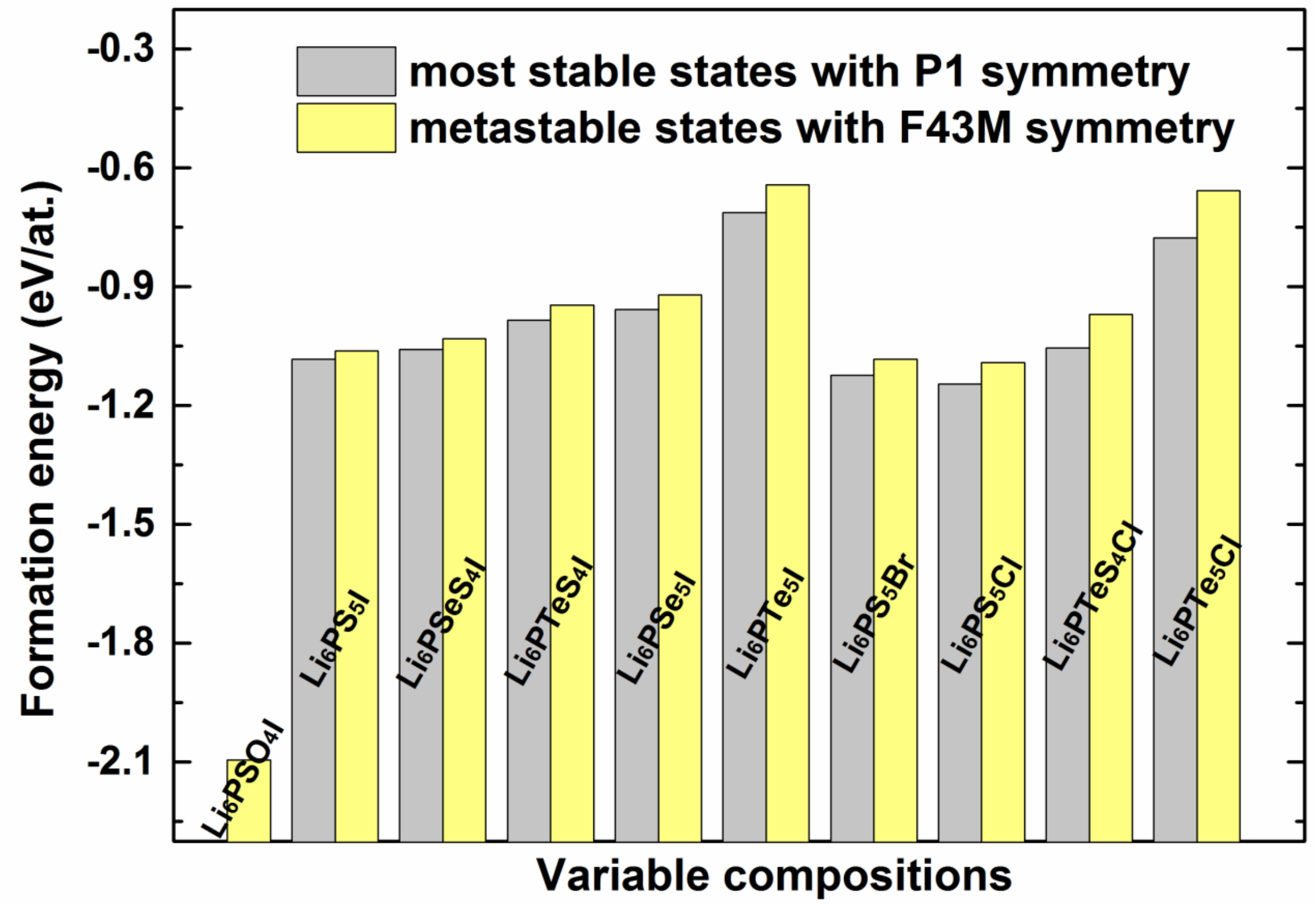

**Fig. 3** Cubic and relaxed structures for (a) $Li_6PS_5I$, (b) $Li_6PSe_5I$, (c) $Li_6PTe_5I$, (d) $Li_6SO_4I$, (e) $Li_6PSeS_4I$, and (f) $Li_6PTeS_4I$. The primitive cells for the stable and cubic phases are compared to reveal the similar basic building blocks.

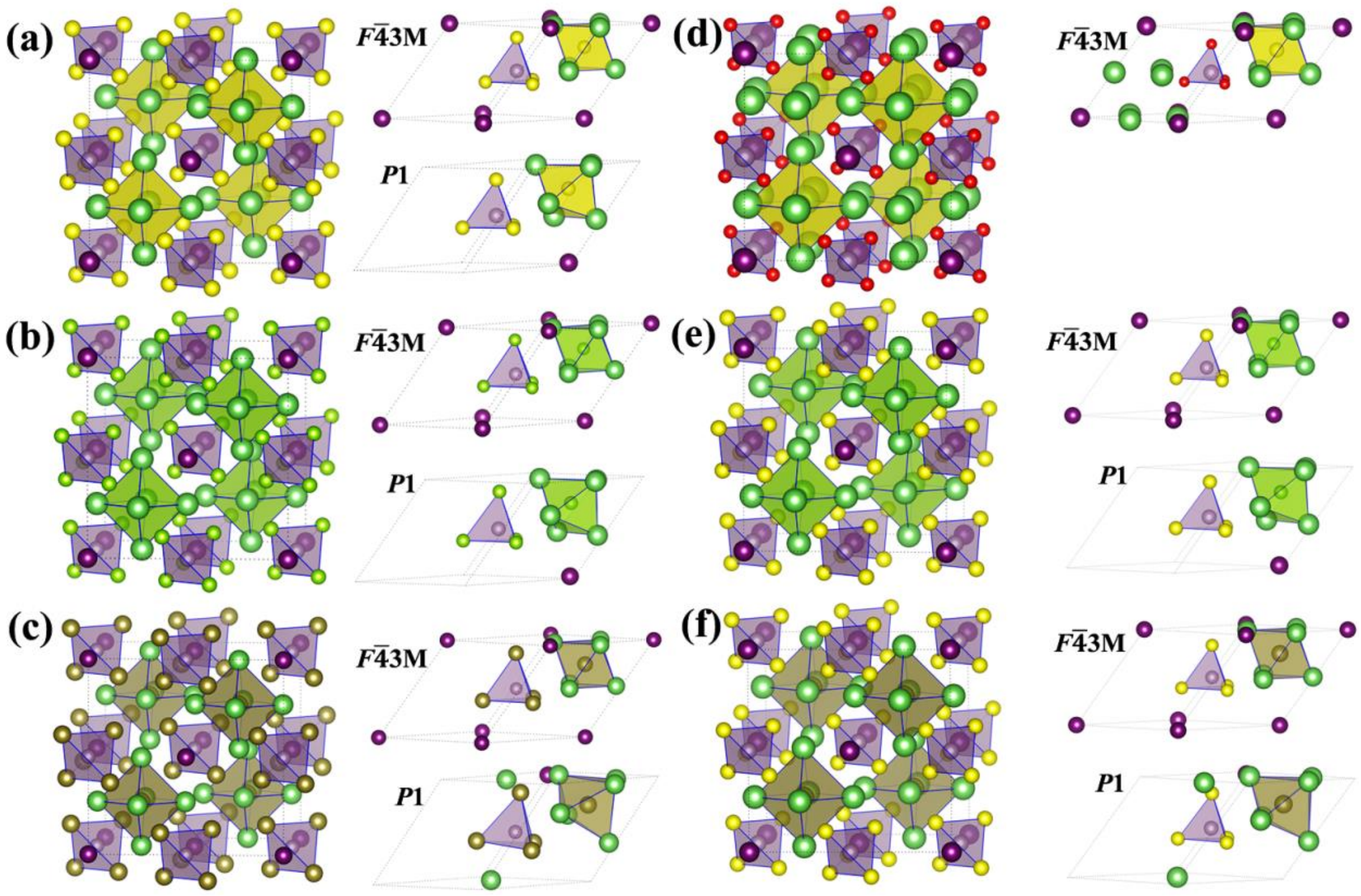

**Fig. 4** Calculated phonon band structures of the stable structures for (a) $Li_6PS_5I$, (b) $Li_6PSe_5I$, (c) $Li_6PTe_5I$, (d) $Li_6PSO_4I$, (e) $Li_6PSeS_4I$, (f) $Li_6PTeS_4I$, (g) $Li_6PS_5Cl$, (h) $Li_6PTeS_4Cl$, and (i) $Li_6PTe_5Cl$. Gray lines define the lower limit for acceptable numerical error, -0.3 THz typically encountered in phonon calculations. Blue lines are the reference state at 0 THz.

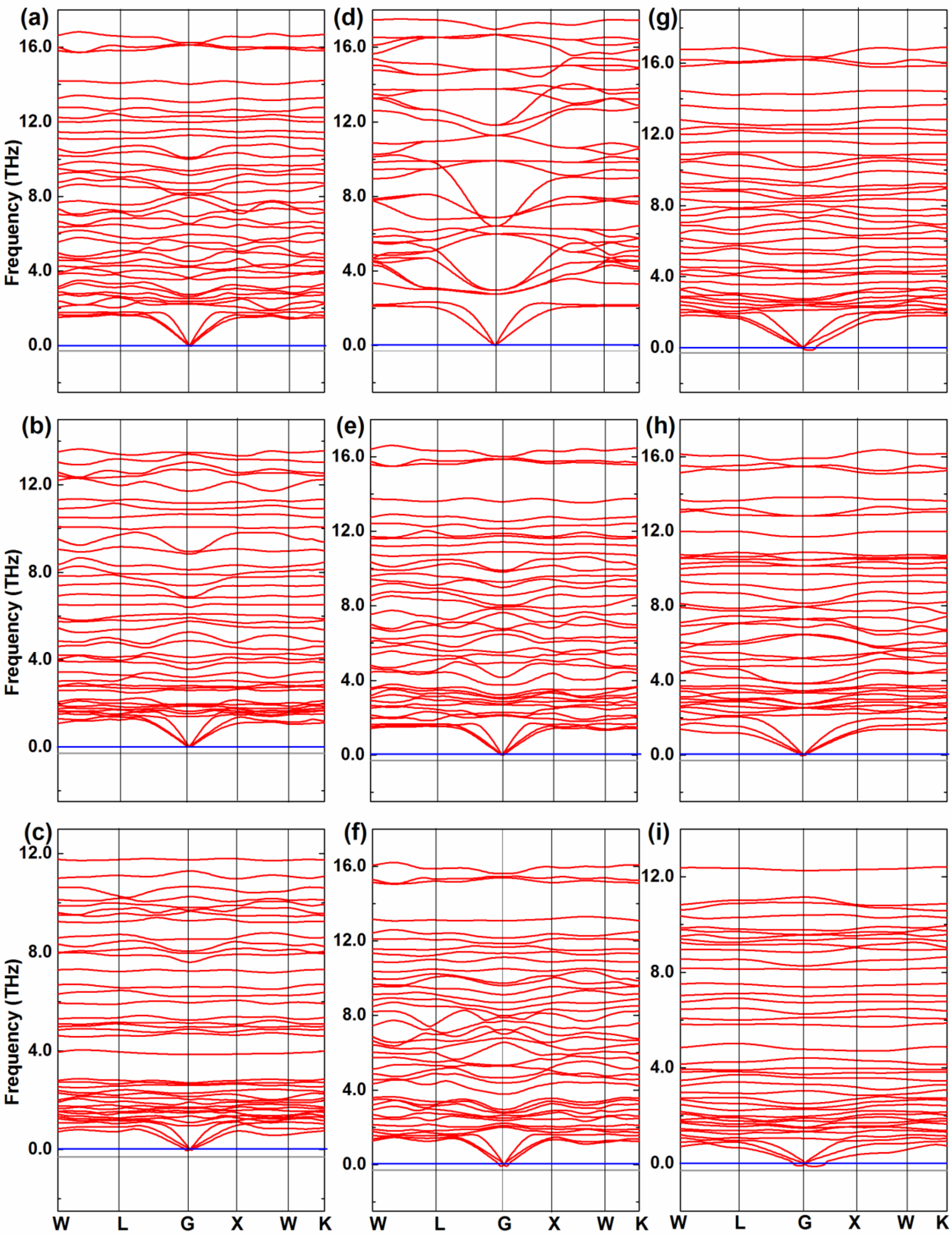

**Fig5** Free energy of formation for the compounds of $Li_6PS_5Cl$ and $Li_6PTeS_4Cl$, with respect to that for constituent stable phases.

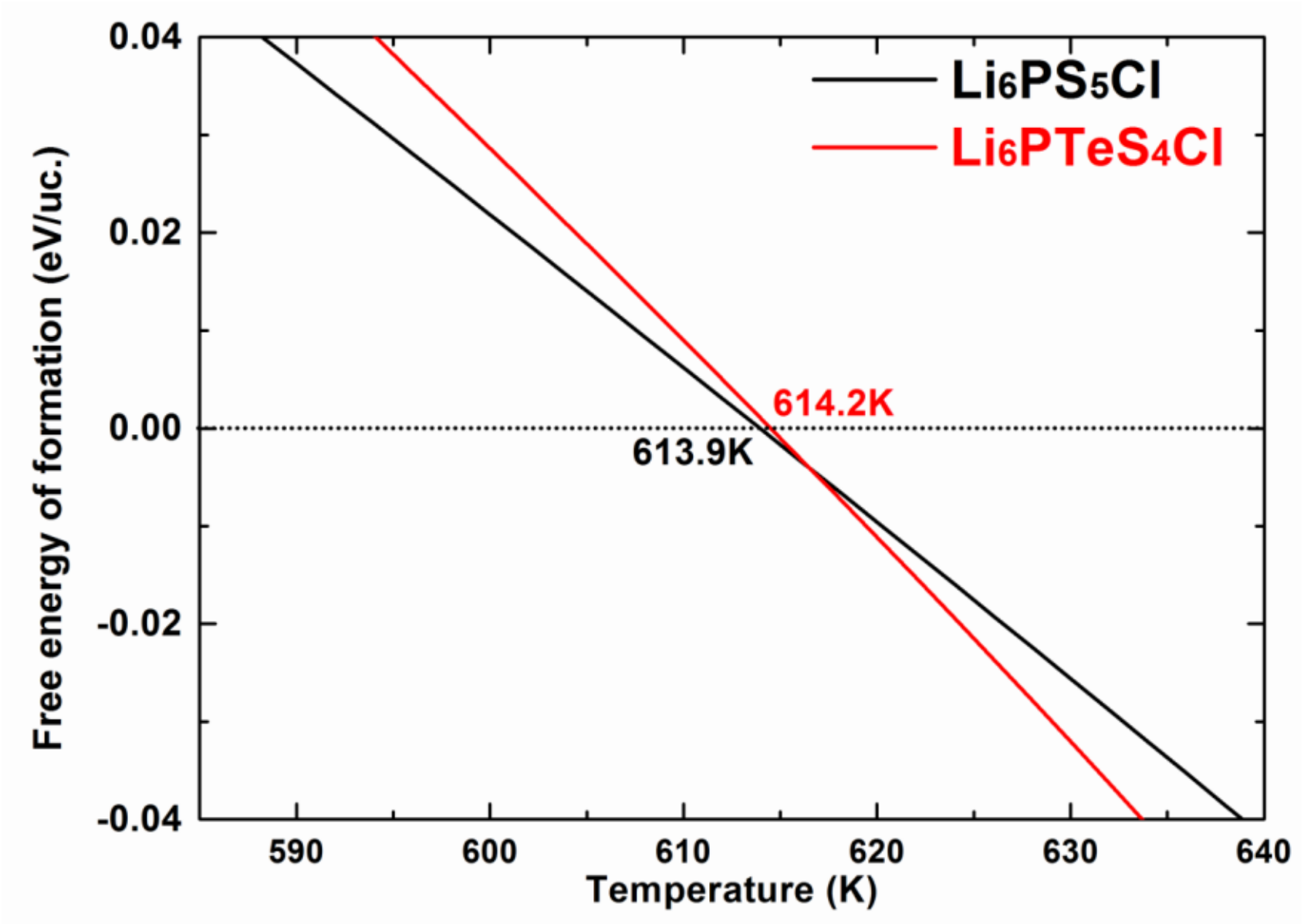

**Fig. 6** Plot of Li uptake per formula unit of solid electrolyte (red solid) against voltage vs Li/Li$^{+}$ for (a) $Li_6PS_5Cl$, and (b) $Li_6PTeS_4Cl$. Text indicates the predicted phase equilibria at corresponding regions at 0 K.

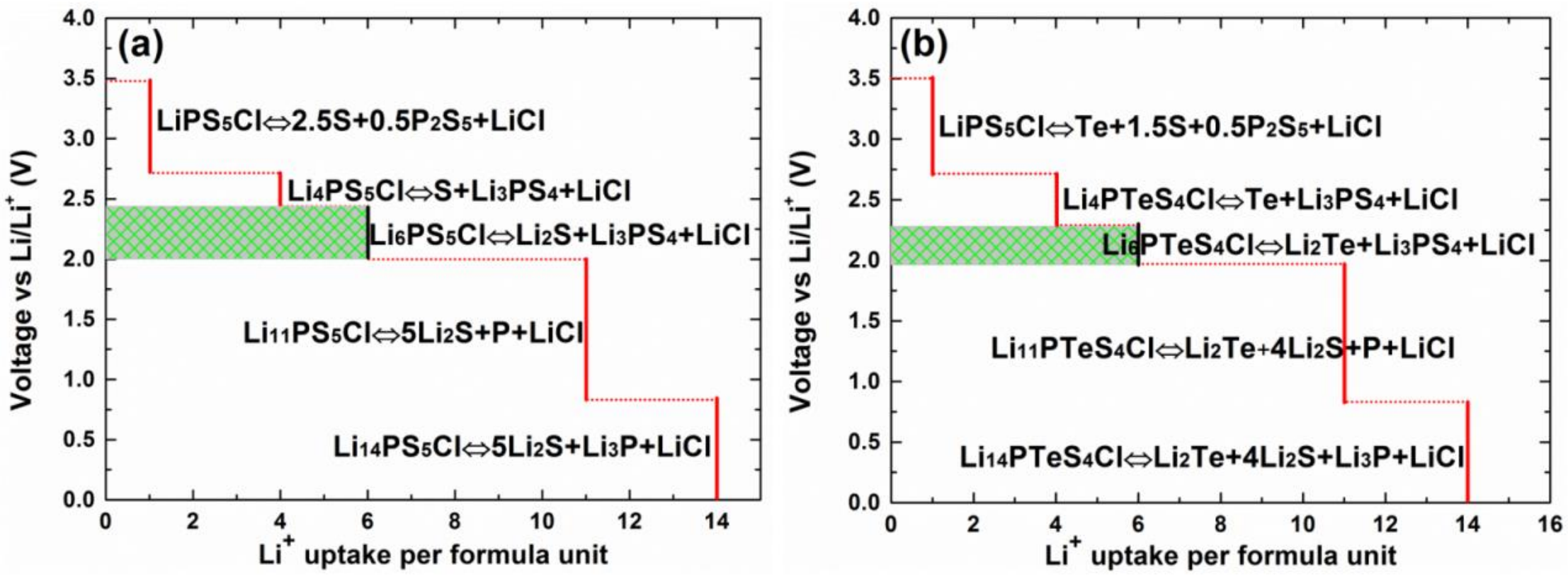

**Fig. 7(a)** Structural model showing the half-way transitional state for inter-octahedral migration of a $Li^+$, sitting within the triangular gate made of one I and two S ions in $Li_6PS_5I$. (b) Calculated energy barriers for intra-octahedral migration of $Li^+$ ions. Calculated inter-octahedral energy barrier in (c) $Li_6PSO_4I$, $Li_6PS_5I$, $Li_6PSeS_4I$, $Li_6PSe_5I$, and $Li_6PTeS_4I$, and in (d) $Li_6PS_5Br$, $Li_6PS_5Cl$, and $Li_6PTeS_4Cl$.

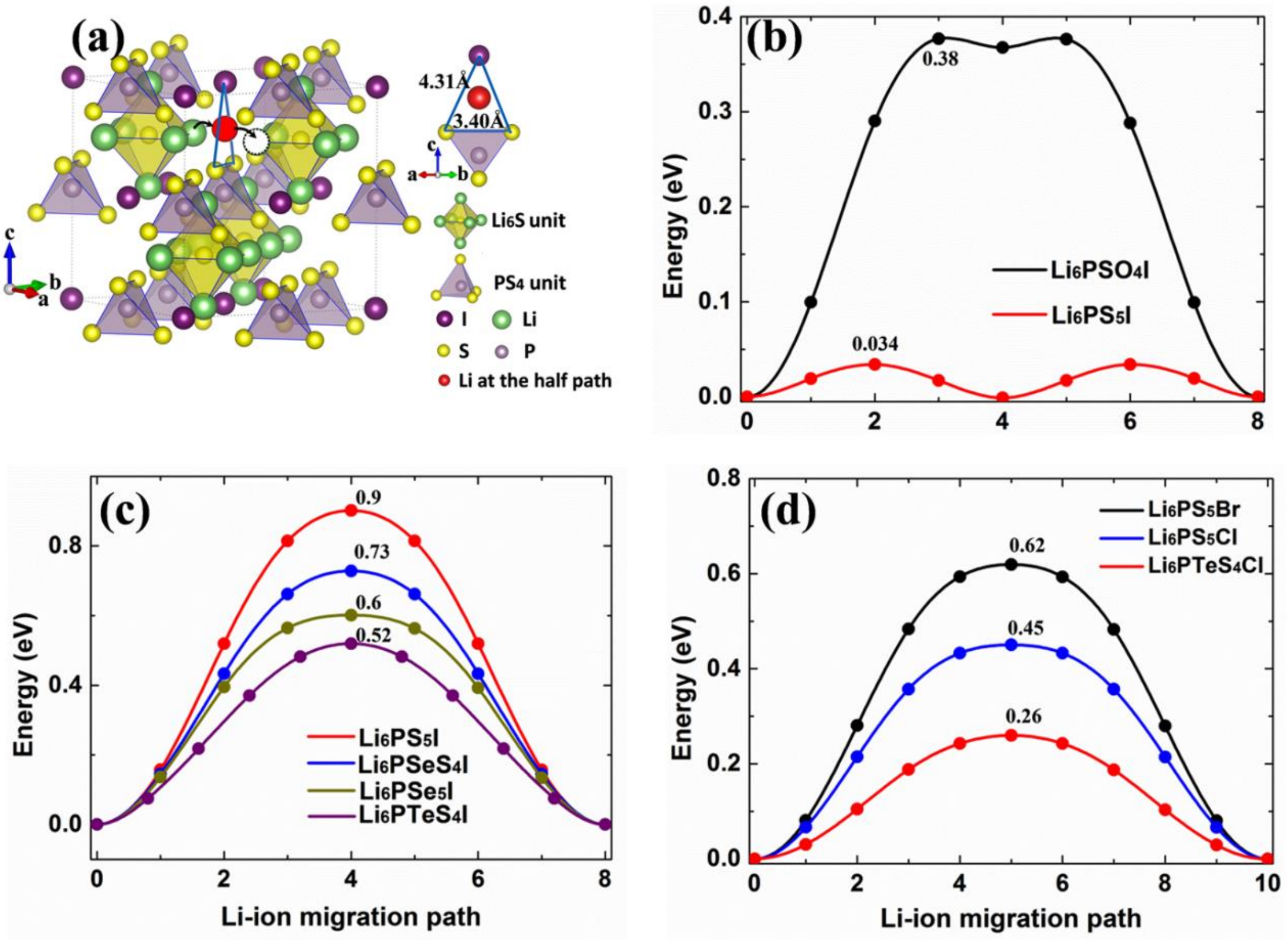

**Fig. 8** Charge maps for (a) cross-section of a $Li_6S$ octahedron in $Li_6PS_5I$, and at the mid-path section in (b) $Li_6PS_5I$, (c) $Li_6PS_5Br$, (d) $Li_6PS_5Cl$, (e) $Li_6PTeS_4I$, and (f) $Li_6PTeS_4Cl$.

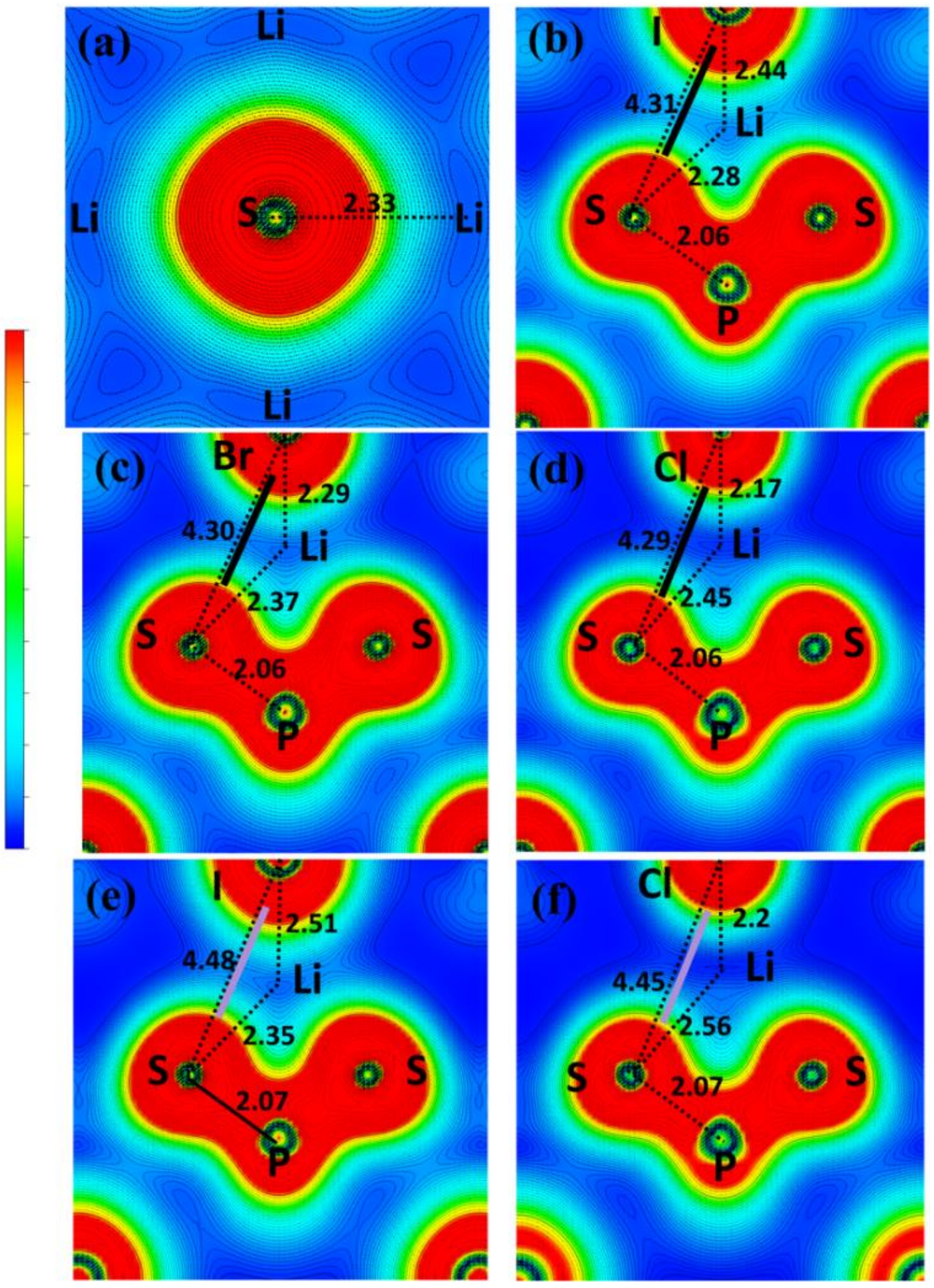

**Fig 9** Arrhenius relationships from the logarithmic plots of D versus 1/T, with activation energies derived from the slops. Literature data[22] for $Li_6PS_5Cl$ based materials are shown as dotted lines for comparison. Replace of S by Te is shown to have significant effect on reduction of diffusion barriers for $Li^+$ ions.

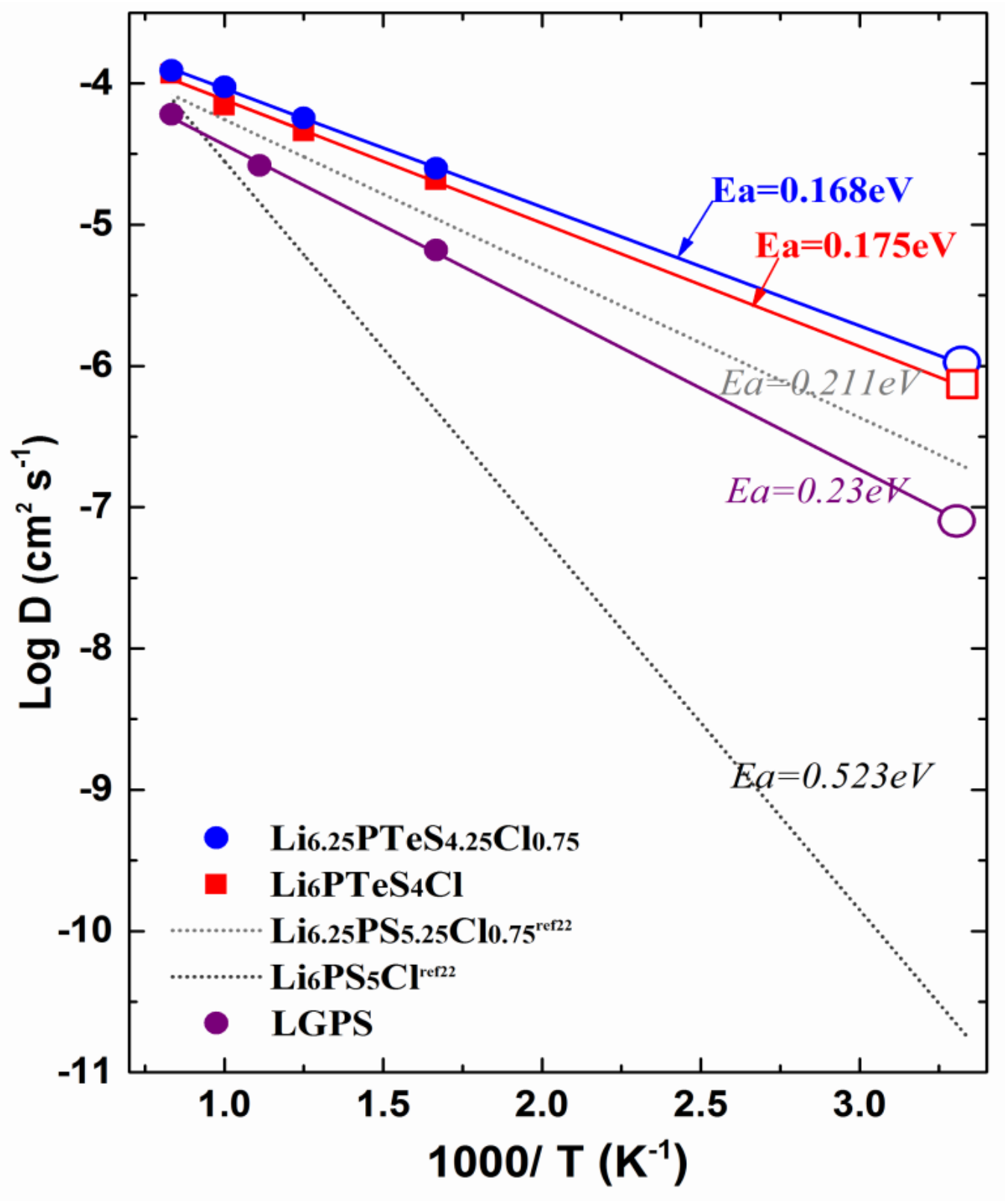

**Fig. 10** The projected density of states calculated using the HSE06 functional: (a) $Li_6PS_5I$, (b), (b) $Li_6PSO_4I$, (c) $Li_6PS_5Cl$, (d) $Li_6PSeS_4I$, (e) $Li_6PTeS_4I$, and (f) $Li_6PTeS_4Cl$.

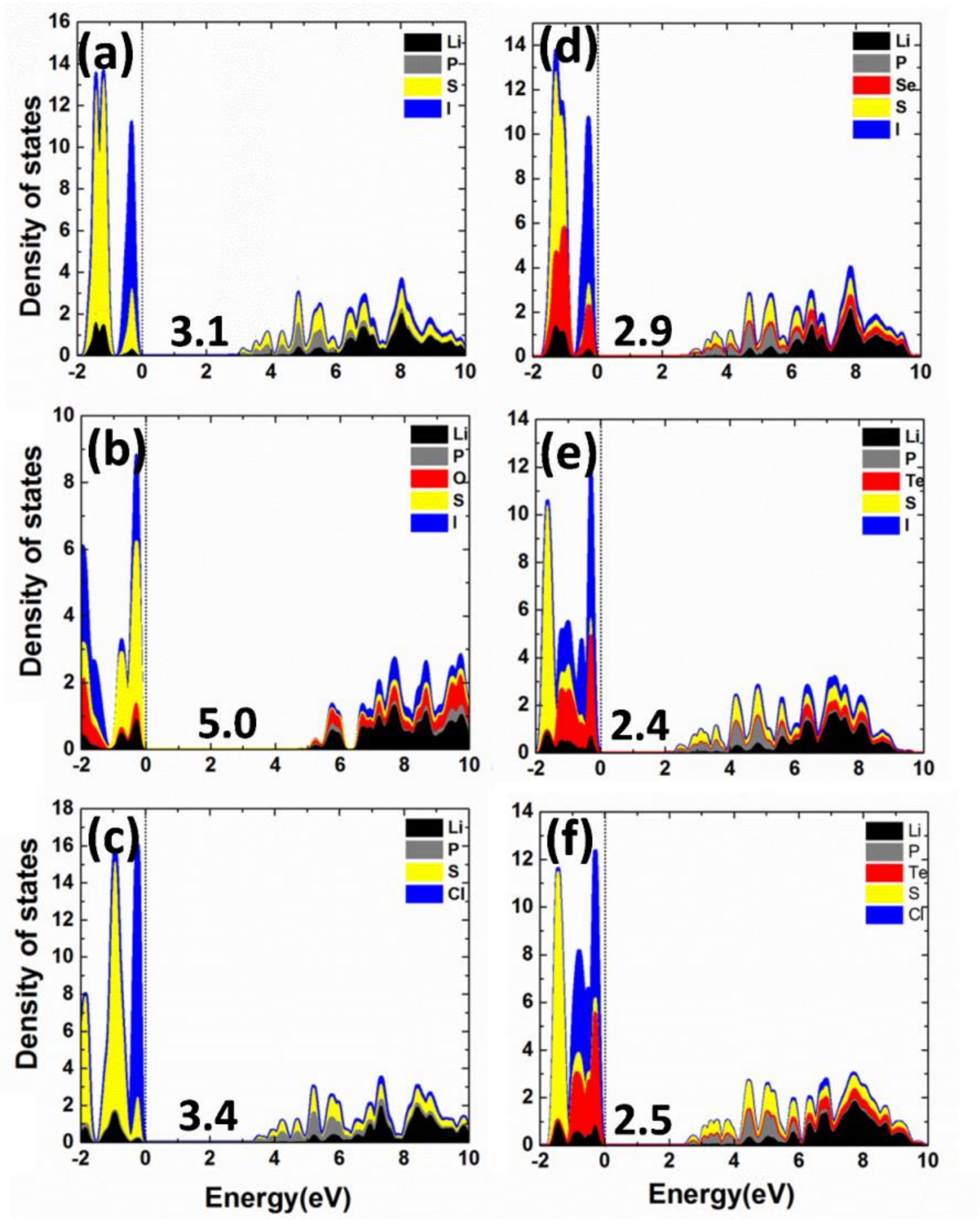